# The Influence of Binary Stars on Dwarf Spheroidal Galaxy Kinematics.

J.C. Hargreaves, G. Gilmore,

*Institute of Astronomy, Madingley Road, Cambridge CB3 0HA.*

J.D. Annan

*105, Sovereigns Quay, Bedford. MK40 1TF*

## Abstract

We have completed a Monte-Carlo simulation to estimate the effect of binary star orbits on the measured velocity dispersion in dwarf spheroidal galaxies. This paper analyses previous attempts at this calculation, and explains the simulations which were performed with mass, period and ellipticity distributions similar to that measured for the solar neighbourhood. The conclusion is that with functions such as these, the contribution of binary stars to the velocity dispersion is small. The distributions are consistent with the percentage of binaries detected by observations, although this is quite dependent on the measuring errors and on the number of years over which measurements have been taken. For binaries to be making a significant contribution to the dispersion measured in dSph galaxies, the distributions of the orbital parameters would need to be very different from those of stars in the solar neighbourhood. In particular more smaller period orbits with higher mass secondaries would be required. The shape of the velocity distribution may help to resolve this issue when more data becomes available. In general, the scenarios producing a larger apparent dispersion have a velocity distribution which deviates more clearly from Gaussian.

# 1  Introduction

A binary star orbits its companion in an elliptical orbit, its velocity changing as it goes. This velocity depends on the masses of the stars, the period of the orbit and the eccentricity of the orbit. The orbit of the stars that we would observe has the centre of mass of the 2 stars as one of the foci. The velocity that we observe at a particular time depends on the position of the star in its orbit and the orientation of the orbit with respect to an observer. If every star observed was on an identical orbit, but at a different point in that orbit, a range of different velocities would be observed, the average velocity being zero. The standard deviation of these velocities is the dispersion produced by this orbit. A particular range of different orbits therefore contributes a specific amount to the velocity dispersion that would be observed. The size of the contribution of these binary stars to the dispersion can be calculated if the number of stars with each set of orbital parameters is known. It is this value that was estimated by the simulation described in this paper.

## 1.1  Previously Published Results

Aaronson & Olszewski (1987) made the first simulations for binary stars in dSph galaxies, taking the mass and period distributions from Galactic studies by Mathieu (1983), and choosing a primary mass of 0.8 $M_\odot$. They chose the phase, inclination angle and eccentricity at random, possibly from uniform distributions, although this is not explicitly stated. They set the binary fraction to 0.5, assumed a certain intrinsic dispersion, and calculated, for each star, the velocities that would be measured from 2 observations taken 1 year apart. They did 500 trials for 10 stars in each case and measured the velocity dispersion after removing variations of greater than 4 km s$^{-1}$ from the sample. They produced no significant deviation from the intrinsic dispersion in their results.

Mateo et al. (1993), Suntzeff et al. (1993), and Vogt et al. (1994) used similar simulations to estimate the effect of binaries in Carina, Sextans and Leo II dSph galaxies respectively. They took the period from a distribution that is uniform in log space, which they state is consistent with various studies, including Duquennoy & Mayor (1991). The mass was taken from a uniform distribution, the primary mass being 0.8 $M_\odot$. The inclination was taken from a cosine distribu-tion, and the eccentricity and phase were chosen at 'random', again, presumably from uniform distributions. The sample sizes for the simulations were similar in size to the observed sample reported in the papers (between 17 and 33 stars), and they repeated the simulation for differing intrinsic velocity dispersions and binary fractions, making 1000 independent trials each time. They calculated the standard deviation and the biweight (Beers et al. 1990) of the resulting velocities. For small intrinsic dispersions of the order of 2 $\mathrm{km\,s^{-1}}$ which is that expected for dSph galaxies containing no dark matter, a binary fraction of 0.2 was required to produce an apparent velocity dispersion of the size observed in dSph galaxies (>6 $\mathrm{km\,s^{-1}}$). They also noticed a difference between the standard deviation and biweight measurements, and suggested that because this was not observed in their observations, it was an indication that binaries did not contribute significantly to the observed dispersion. This would require the binary star fraction in dSph galaxies to be considerably less than that observed in the solar neighbourhood.

## 1.2 Observations of Binary Stars

Hargreaves et al. (1994a, 1994b) have made multi-epoch observations of 18 stars in Sextans and Ursa Minor. Of these, 2 stars show velocity variations which indicate that they may be binary stars. The variation of 1 of these stars is a far more significant detection than the other. If both these stars are binary stars the observed binary fraction was 0.11 over 2 years of observations. It is important to know what actual binary fraction this represents, since if less than half of the binaries were observed in 2 years this would imply a binary fraction of greater than 0.2. Other observations of stars in dSph galaxies have obtained between 4 and 12 years of repeat measurements of 63 stars in Sculptor, Fornax, Ursa Minor and Draco. These observations have found a binary fraction of between 0.1 and 0.16 (Mateo, 1994). For Draco where there are 24 stars for which there are up to 5 observations of each, 4 appear to be binaries (Mateo, 1994). Our simulation was designed to place an estimate on the fraction of binary stars that would be observed over a certain period of observations, as well as to calculate the contribution binary orbits make to the measured velocity dispersion.

# 2   Details of the Model.

The model was constructed to examine the velocity distribution caused by the orbits of binary stars. The model made a Monte-Carlo simulation, choosing binary orbits with parameters chosen at random from empirical and theoretical and distributions for a large number of stars, and then evolved each star round its orbit. From this it was possible to ascertain what fraction of binary stars would be identified over the course of a certain number of equally spaced observations, given a certain velocity above which a velocity difference would become apparent to an observer. The velocity distribution obtained from evolving the stars round their orbits could be used to calculate the velocity dispersion caused by the binary stars.

The model chose the parameters of the binary orbits randomly from different distributions. The velocity that would be measured by an observer was calculated for equal time intervals all the way round each orbit. In this way, a distribution of velocities was obtained for each orbit. Of these velocities 100 were chosen at random from the set of velocities for each star and written to file 1, ensuring that the distribution for each orbit represented in the file had equal weight. Thus, file 1 contained the velocity distribution for the set of binary orbits.

It was assumed that a specific difference in velocity, called the threshold velocity, could be detected between 2 velocity measurements made by the observer: the value depended on the assumed measuring errors. If the difference in velocity between time intervals, measured along the line of sight, was equal to or greater than this value, then the velocities concerned were marked. All the velocities that appeared in file 1 and were not marked in this way were written to file 2. Therefore file 2 contained the velocity distribution for the binary orbits which would not be identified by the assumed measuring errors. The fraction of the binary stars that would be identified was given by the ratio of the number of velocities in files 2 and 1. The velocity dispersion of the binary distribution ($\sigma_b$ in equation 12) was equal to the standard deviation of the velocities in file 1, whereas the velocity dispersion that would be obtained if the stars were thrown out of the sample when identified as belonging to binary systems was the standard deviation of the stars in file 2. The standard deviations of the data in files 1 and 2 were consistent under repetition for given parameter distributions, provided that a sufficient number of stars were used in the sample.

The orbital parameters which need to be considered are: the masses of the 2 stars; the period of the orbit; the minimum approach distance of the stars; the ellipticity of the orbit; the inclination; the phase; and the position of the apocentre with respect to the observer.

## 2.1 The Distributions of the Orbital Parameters

The best estimate for the distributions of the orbits of binary stars comes from the solar neighbourhood sample observed by Duquennoy & Mayor (1991), so it is these distributions that have been used to define the orbital parameters. The primary stars in the orbits of that study were solar mass G dwarfs. Although the masses of the stars observed in the dSph galaxies are fairly close to the masses of these dwarfs, it may well be the case that the orbital distributions discovered are not applicable in the very different conditions of a dSph galaxy. The simulations which used these orbit distributions, assuming the radius of the primary to be 10 and 30 $R_\odot$, are termed DM10 and DM30 respectively in the rest of this paper. To compare this with previous simulations, the same distributions used by Mateo et al. in their simulations were also used (hereafter Ma). The secondary mass distribution calculated by Kroupa, Tout & Gilmore (1993), which rises rather than falls towards low masses, was also used. In these simulations, the DM distributions were used for the other parameters apart from mass: the simulations are called KTG10 and KTG30.

### 2.1.1 Mass

Carbon stars are the brightest stars in dSph galaxies, but these are few in number and have a high probability of being velocity variables (McClure 1984). The next brightest stars, occupying the tip of the giant branch, are the K giants and it is observations of these that are used to calculate the velocity dispersions. Given the stellar populations in dSph galaxies are predominantly of intermediate to old age, these stars will have mass of about 0.8 $M_\odot$, so this is the value of the primary in our simulations. Figure 1 shows that there is very little difference in the results of the simulation taking a primary mass of between 0.6 and 1.0 $M_\odot$.

The secondary mass distribution found by Duquennoy & Mayor is given by

$$\mathbf{P}(M2) \propto \exp\left(\frac{-(M2-0.23)}{0.42}\right)^2, \qquad (1)$$

where $M2$ is allowed to vary between 0.05 $M_\odot$ and the mass of the primary.

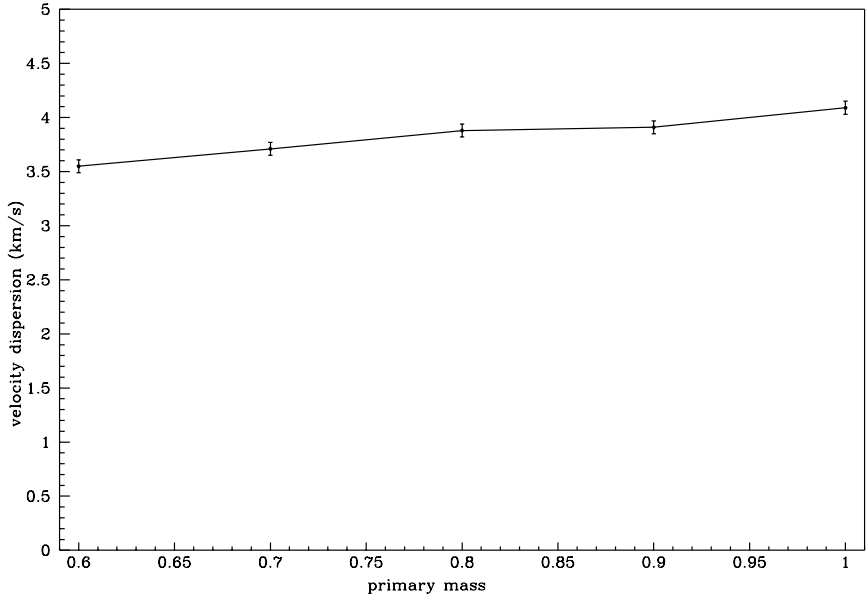

Figure 1: The velocity dispersion obtained for different primary masses for model DM10.

Mateo et al. also took the primary mass to be 0.8 $M_\odot$, but their secondary masses were taken from a uniform distribution with masses varying between 0.05 and 0.8 $M_\odot$.

Kroupa, Tout & Gilmore's mass distribution has the following form:

$$\mathbf{P}(m) = \begin{cases} 0.035 m^{-1.3} & 0.08 \leq m \leq 0.5 \\ 0.019 m^{-2.2} & 1.0 \leq m \leq 1.0 \\ 0.019 m^{-2.7} & 1.0 \leq m < \infty. \end{cases} \qquad (2)$$

### 2.1.2 Period

The Duquennoy & Mayor period distribution is consistent with the estimate of Kroupa, Tout & Gilmore (1990). It is Gaussian in log space, and has the form

$$\mathbf{P}(\log(P_{days})) \propto \exp\left(\frac{-(x-4.8)}{2.3}\right)^2 \qquad (3)$$

where $P_{days}$ is the period in days. No maximum or minimum bounds were imposed on this distribution.

The distribution used by Mateo et al. was uniform in the logarithm of the period. They stated that this distribution is also consistent with the results of Duquennoy & Mayor.

### 2.1.3 Ellipticity

Here, the ellipticity, e, is defined to be $e = (1 - b/a)$, where $b$ and $a$ are the minor and major axes of the orbit respectively.

Duquennoy & Mayor found the ellipticity obeyed the following distributions.

$$\begin{array}{ll} \text{period} < 11 \text{ days} & \mathbf{P}(e) = 0.0 \\ 11 \text{ days} < \text{period} < 1000 \text{days} & \mathbf{P}(e) \propto \exp\left(\frac{-(x-0.3)}{0.16}\right)^2 \\ \text{period} > 1000 \text{days} & e \propto 2e \end{array} \qquad (4)$$

Mateo et al. took the ellipticity from a uniform distribution in the range 0.5 to 1.0.

### 2.1.4 Angles

Figure 2 shows the orbit of a binary star round its centre of mass. Viewed from the earth, the star moves in an ellipse in the plane perpendicular to our line of sight. The inclination, $i$, is the angle between the orbital plane and the viewing plane. For spherical symmetry, the normal to the viewing plane must be evenly distributed over the sphere, implying that the distribution of the orbital inclinations is proportional to the sine of the inclination.

$$\mathbf{P}(i) \propto \sin i. \qquad (5)$$

The angle $w$ is the angle between the ascending node, $M$, and the apastron of the orbit. As in the diagram, the phase, $v$, is the angle between $w$ and the current position of the star, taken in the direction shown. Both $w$ and $v$ were allowed to vary between 0 and 360 degrees, so, because of symmetry, $i$ was allowed to vary between 0 and 180 degrees. The angle $w$ was chosen from a uniform distribution, and was a constant parameter of the orbit. The phase, $v$, however is the angle that varies with time, so that although its initial value was chosen from a uniform distribution, its value thereafter varied in accordance with Kepler's 2nd law.

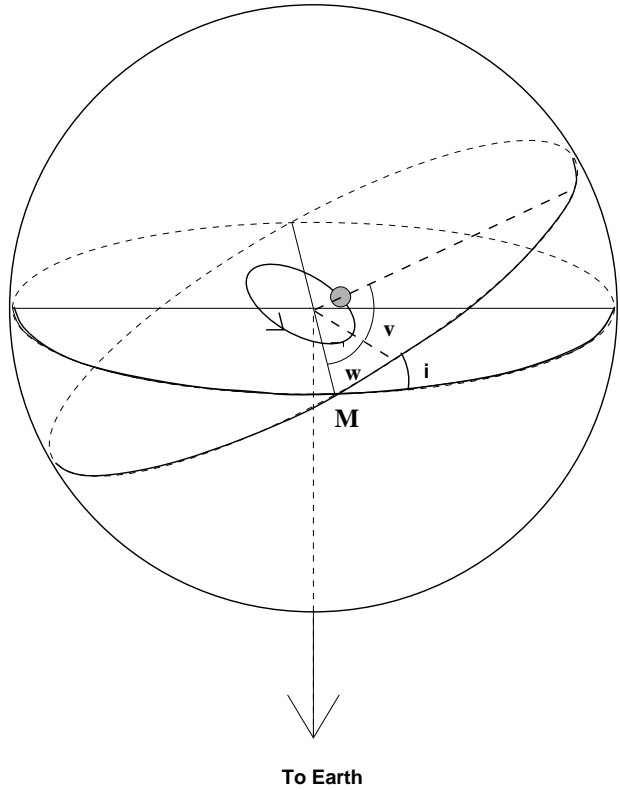

Figure 2: Orbit of a binary star round the centre of mass.

## 2.2 Radius Cutoff

Orbits of very low period or high ellipticity may not be physically possible, since they can result in too close an encounter between the 2 stars in the binary. To estimate this minimum distance a simple gravitational Roche-Lobe radius estimate was used.

$$M2 = M1 \times \frac{a_p^2(a_p - R)}{R^3(2a_p - R)}. \qquad (6)$$

Here $a_p$ is the distance between the stars at apastron, and $R$ is the radius of the primary star of mass $M1$, $M2$ being the maximum mass of the secondary before

Roche-Lobe overflow occurs. The radius of the giant stars in dSph galaxies is not well known, as the evolution of stars in low metallicity environments such as that found in dSph galaxies is not understood in detail, so a range of between 10 and 30 $R_\odot$ was used in the simulations, which should cover the possibilities. This is a difference between the binaries in dSph galaxies and the binaries with solar mass primaries observed by Duquennoy & Mayor in the solar neighbourhood: some of the orbits which existed round the primaries when they were main sequence stars should no longer exist, so it may be expected for the fraction of binary stars to be somewhat lower than that observed in the solar neighbourhood.

## 2.3 Velocity

From the period, the semi-major axis of the real orbit, rather than that with respect to the centre of mass, is given by

$$a^3 = (M1 + M2)T^2 \qquad (7)$$

where $a$ is the semi-major axis in astronomical units, $T$ is the period in years and $M1$ and $M2$ are the masses in solar masses.

The line of sight velocity observed for a particular binary star with mass M1 at some phase, $v$ is given by

$$V = 2\pi \frac{1.49598 \times 10^{11}}{365.25 \times 3600} \frac{M2 \sin i}{\sqrt{a(M1+M2)}} \left[ \frac{\cos(v+w) + e\cos w}{\sqrt{(1-e^2)}} \right]. \qquad (8)$$

Here the velocity is in ms$^{-1}$, and the semi-major axis, $a$, is in astronomical units. The ellipticity, $e$ is $\sqrt{1-(b/a)^2}$.

## 2.4 Time

To calculate the velocity that would be observed at equal time intervals round the orbit using equation 8, it was necessary to calculate the change in the phase resulting from a change in time.

Figure 3 shows the geometry of the orbit. The secondary, M2 sits at the focus of the ellipse orbited by M1. The phase, $v$, is also called the true anomaly and $E$, is the eccentric anomaly. Geometry and Kepler's 2nd law can be used to derive the following relationships between these 2 angles and time.

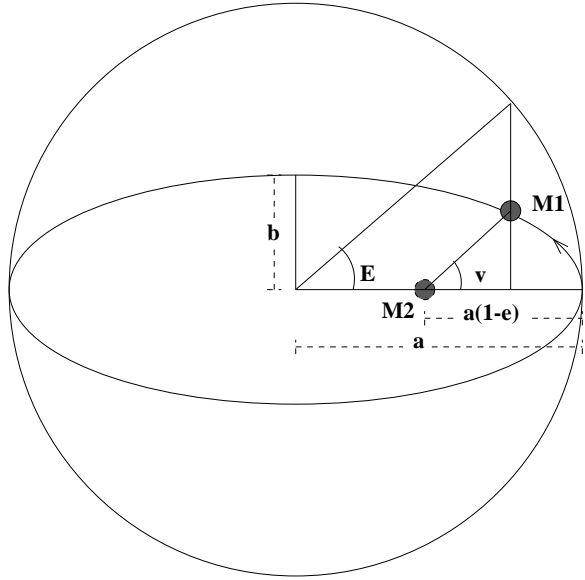

Figure 3: Geometry of the orbit of 2 binary stars

$$\tan \frac{E}{2} = \sqrt{\frac{(1-e)}{(1+e)}} \tan \frac{v}{2}. \tag{9}$$

$$E - \sin E = \frac{2(\tau)}{T}. \tag{10}$$

Here $\tau$ is the time since the stars were last at apastron.

So we obtain $E$ from $v$ using equation 9, iterate to find the new $E$ from equation 10 and then use equation 9 once more to calculate the new $v$. In this way and using equation 8 the velocity can be calculated at every stage of the orbit.

## 2.5 Fraction of Binary Stars Identified

There were some binary stars that would never be identified because the range of their orbital velocities was small compared to the measuring error. These were the stars for which the difference between the maximum and minimum value of the velocity from equation 8 was less than the threshold velocity at which the difference would appear significant to the observer (this is explained for specific examples in Section 3). The maximum and minimum values were obtained by evaluating the equation at $v = -w$ and $v = \pi - w$. For these stars, the orbit was divided into 100 equal timesteps and each velocity produced was written to files 1 and 2.

The time difference between measurements in the simulations was taken to be 1 year. Therefore each star not already catered for by the criterion of the previous paragraph was evolved round its orbit, recording the velocity at 1 yearly time intervals. For periods of less than 100 years 1001 yearly velocities were taken. For periods greater than this the number of orbital revolutions was reduced to save computing time. For periods between 100 and 1000 years velocities were calculated 5 times round the orbit, and for those between 1000 and 10000 years 2 orbits' worth of velocities were calculated. Then all the differences between consecutive velocities were calculated, thus obtaining 1000 velocity differences for periods of less than 100 years.

The velocities were marked for which, if observations were started from the position on the orbit associated with that velocity, over the course of the observations, the star would be identified as having a binary orbit. Therefore another parameter was used, which was the number of years over which observations were conducted. A velocity was marked if any of the subsequently obtained velocity differences, or their sum, was greater than the threshold velocity. Then 100 velocities were chosen at random with equal probability from the sample for each star, writing all 100 to file 1 and only those not marked to file 2.

For periods greater than 10000 years a slightly different approach was adopted. For such large periods the only portion of the orbit at which the star might possibly be identified as a binary was close to apastron, so 1001 velocities were calculated symmetrically about apastron, and the phase angles about which the velocity differences were sufficiently large to be detected over the years of observation were calculated. Then the velocity was calculated at 100 equal timesteps right round the orbit (ie each timestep took went 100th of the way round the orbit) and marked those which fell between the calculated phase points. These velocities were written to the files in the same way as the velocities for the other orbits.

For the periods of less than 10000 years, the probability of recognising a particular binary star orbit over the observing period was equal to the number of marked velocities, divided by the number of observations in total. For the larger periods, the probability was equal to the number of velocities within the calculated phase range where the binary could be identified, divided by the number of timesteps contained in an orbit. For a timestep of 1 year, this denominator is just the orbital period in years.

The average of the probabilities for all the stars is equal to the fraction of stars that would be recognised as having binary orbits.

## 3  Results

Figures 4, 5, 6, 7 and 8 show the results for the simulations. Each figure contains 2 plots. The upper plot shows the percentage of the binaries that would be identified for a certain threshold velocity. The different lines are for yearly observations spanning 2, 10 and 50 years. The solid line on the lower plot shows the standard deviation of the distribution arising from binary stars with the orbital parameters chosen: the dotted lines show the standard deviation of the distributions once the identified binary stars have been removed.

For observations with an error for each velocity measurement of $\sigma_{err}$, the threshold velocity for a binary star to be detected was $3\sqrt{2}\sigma_{\rm err}$. For the observations of Hargreaves et al. (1994a, 1994b) the error per velocity was close to 2 km s$^{-1}$. Aaronson & Olszewski (1987), took 4 km s$^{-1}$ as the threshold velocity in their simulation, whereas the average error on the velocities of Pryor, Olszewski & Armandroff (1995) was 3.6 km s$^{-1}$, giving a threshold velocity of 15.3 km s$^{-1}$. Threshold velocities of 1, 4, 8.5, 15.3, and 21.2 km s$^{-1}$ were used, as shown in the plots, to cover the range of possible observations.

The simulation was made using 10000 stars since it was at this level that repeat simulations became easily recognisable. The one standard deviation of the percentage of identified binaries taken from repeat simulations was about 0.5%, whereas that for the velocity dispersion was about 0.06 km s$^{-1}$. Table 1 shows the standard deviation of the binary star distribution required for binary stars to be making a significant contribution to the measured dispersion in dSph galaxies; dispersions due to binary orbits alone are required to be greater than about 6 km s$^{-1}$ to explain the observed dispersion assuming a low mass-to-light ratio. The results

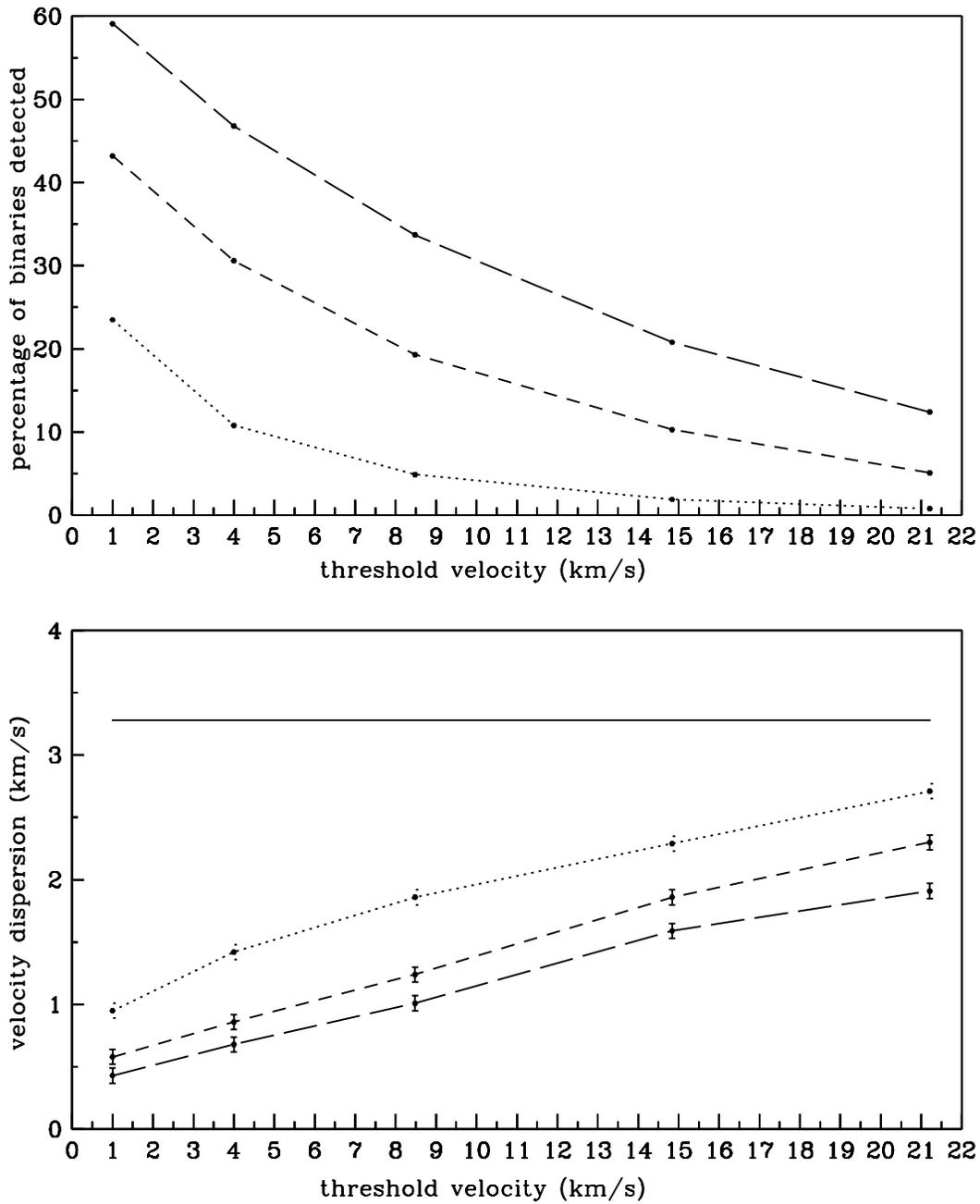

Figure 4: Model Ma. The solid line in the bottom plot shows the velocity dispersion of the whole sample. The dashed and dotted lines show the percentage of binaries that would be detected (top plot) and the velocity dispersion of the residual sample (bottom plot) after 2 years of observations. The short dashes and long dashes show the same statistics for 10 and 20 years of observations respectively

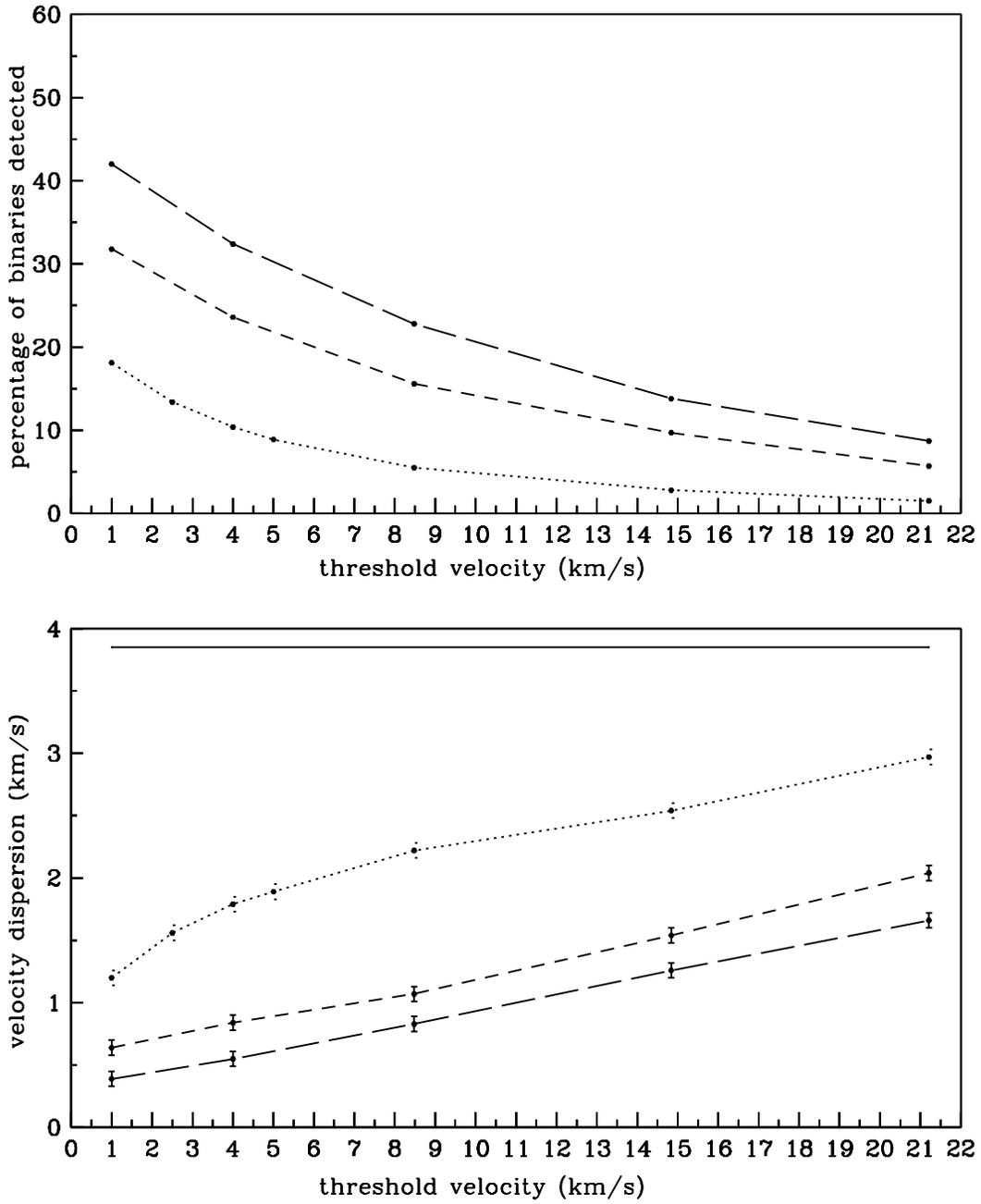

Figure 5: Model DM10. The solid line in the bottom plot shows the velocity dispersion of the whole sample. The dasked and dotted lines show the percentage of binaries that would be detected (top plot) and the velocity dispersion of the residual sample (bottom plot) after 2 years of observations. The short dashes and long dashes show the same statistics for 10 and 20 years of observations respectively

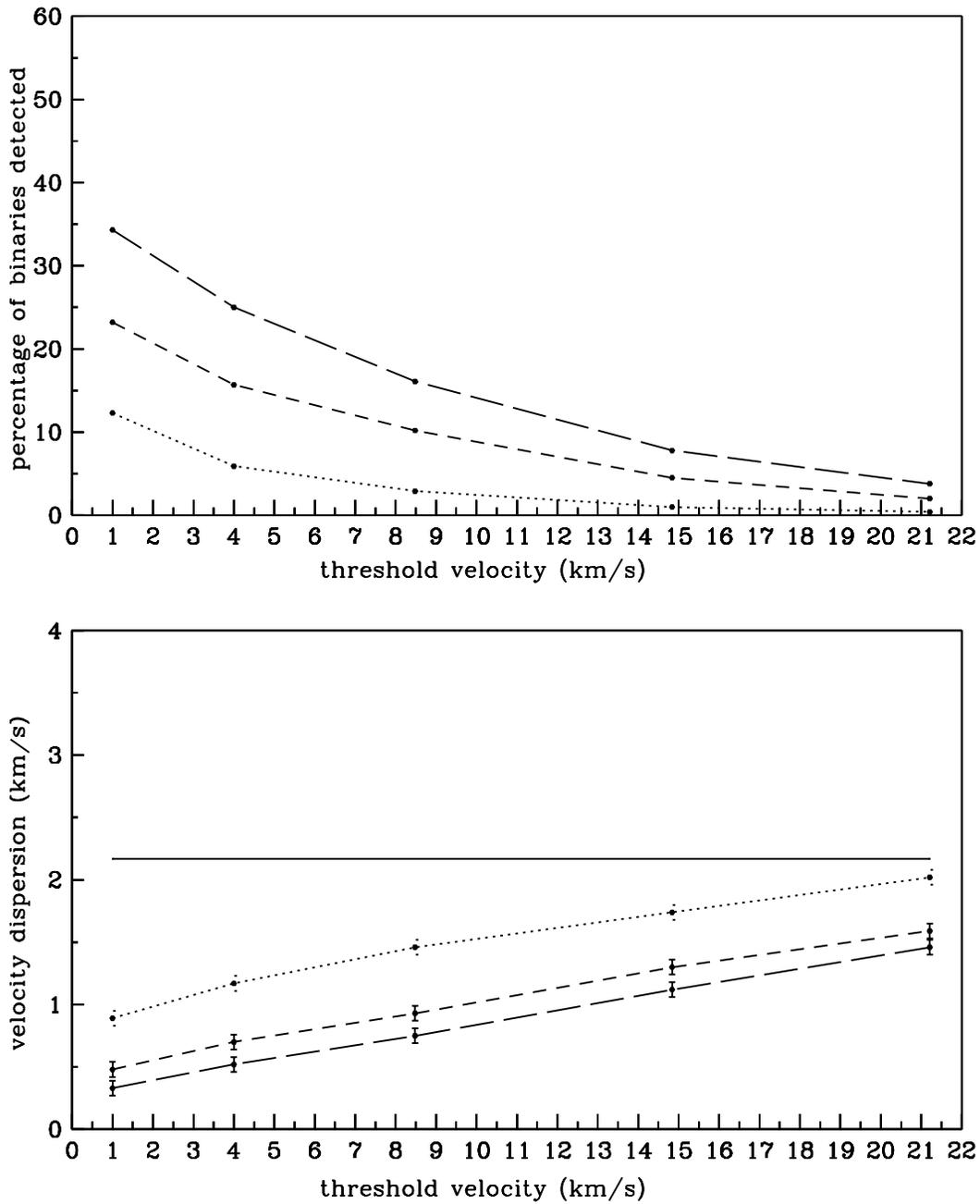

Figure 6: Model DM30. The solid line in the bottom plot shows the velocity dispersion of the whole sample. The dashed and dotted lines show the percentage of binaries that would be detected (top plot) and the velocity dispersion of the residual sample (bottom plot) after 2 years of observations. The short dashes and long dashes show the same statistics for 10 and 20 years of observations respectively

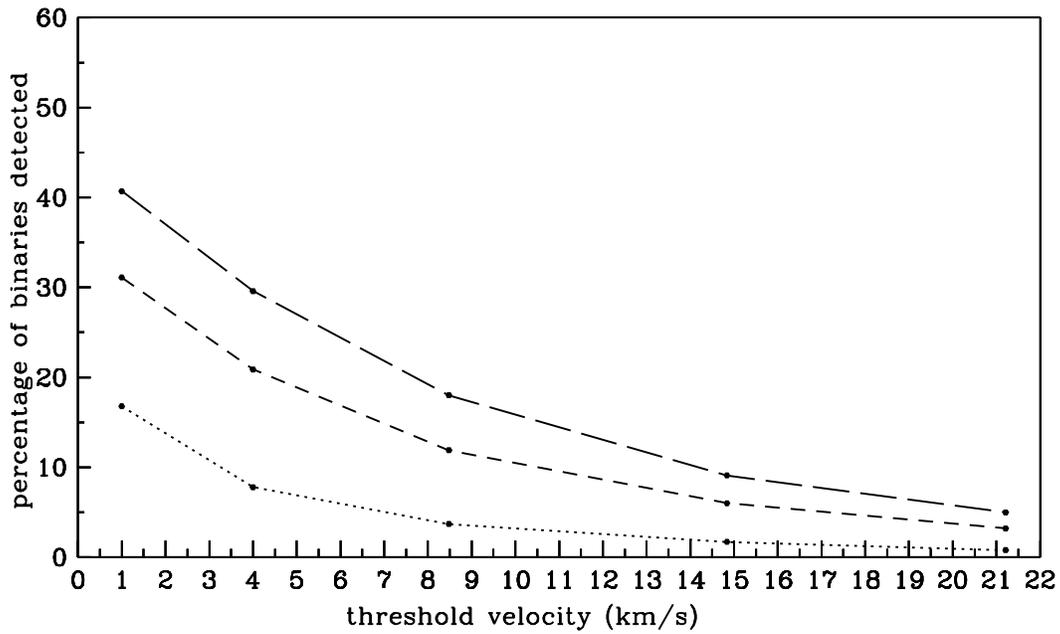

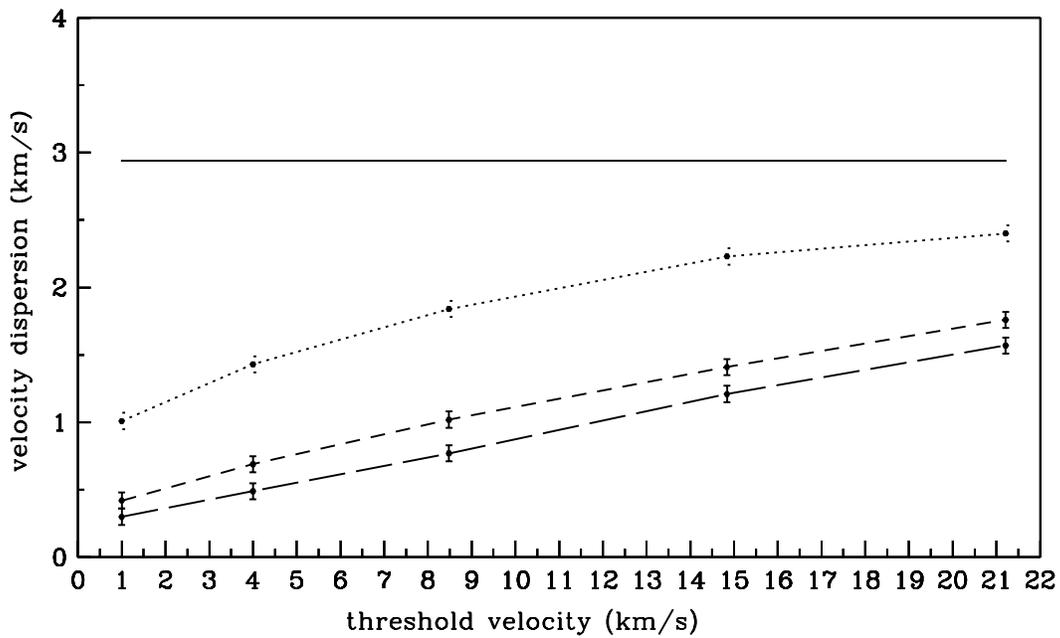

Figure 7: Model KTG10. The solid line in the bottom plot shows the velocity dispersion of the whole sample. The dashed and dotted lines show the percentage of binaries that would be detected (top plot) and the velocity dispersion of the residual sample (bottom plot) after 2 years of observations. The short dashes and long dashes show the same statistics for 10 and 20 years of observations respectively

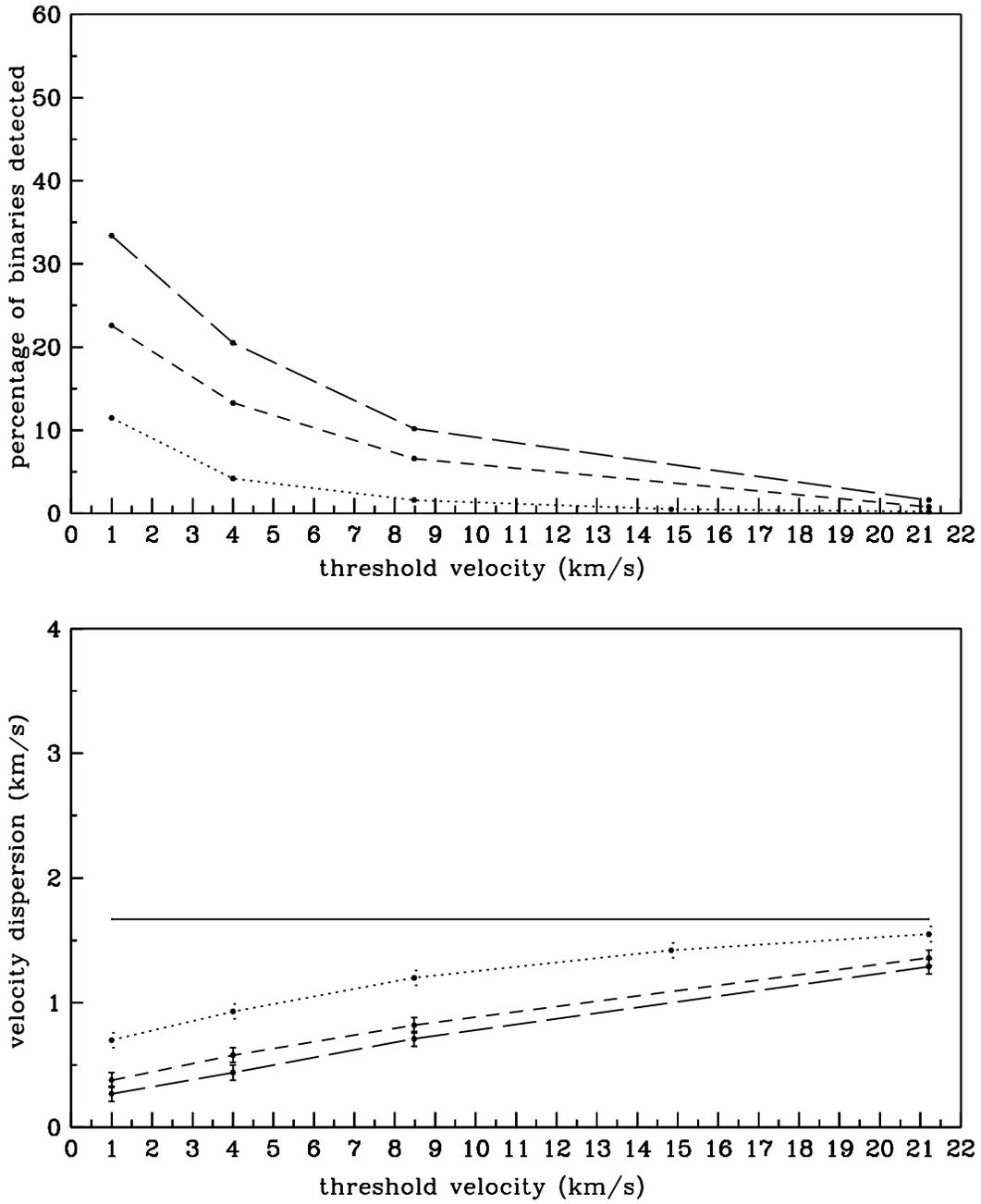

Figure 8: Model KTG30. The solid line in the bottom plot shows the velocity dispersion of the whole sample. The dashed and dotted lines show the percentage of binaries that would be detected (top plot) and the velocity dispersion of the residual sample (bottom plot) after 2 years of observations. The short dashes and long dashes show the same statistics for 10 and 20 years of observations respectively

Table 1: Velocity dispersion required from the binary stars for binary orbits to account for the excess dispersion over that generated for a typical stellar mass-to-light ratio. All velocities are in $\mathrm{km\,s^{-1}}$.

Measured dispersion = 7 $\mathrm{km\,s^{-1}}$

| binary fraction | intrinsic dispersion 2 | 4 |
|---|---|---|
| 0.25 | 13.4 | 11.5 |
| 0.5 | 9.5 | 8.1 |
| 0.75 | 7.8 | 6.6 |
| 1.0 | 6.7 | 5.7 |

Measured dispersion = 10 $\mathrm{km\,s^{-1}}$

| binary fraction | intrinsic dispersion 2 | 4 |
|---|---|---|
| 0.25 | 19.6 | 18.3 |
| 0.5 | 13.9 | 13.0 |
| 0.75 | 11.3 | 10.6 |
| 1.0 | 9.8 | 9.2 |

from the simulations from the chosen paramter range show that the highest dispersion caused by the binary orbits alone is about 3 $\mathrm{km\,s^{-1}}$, We conclude that either the velocity dispersions that have been observed are largely unaffected by binary stars, or that the orbital parameters, which were after all taken from Galactic observations, are inappropriate for dSph galaxies.

Comparing the results from the Ma, DM10 and DM30 models, it can be seen that the percentage of binaries detected in model Ma was considerably larger than that in the other 2 cases. This is largely due to the fact that the orbits in the Ma model have a maximum period of 10000 years. The periods in the DM model have no such upper bound and have about 30% of periods above 10000 years. These long periods alone produce only a small dispersion, and as they can only be detected during the few years close to apastron, they only slightly increase the total number of binary stars identified. When an extra 30% of stars are added analytically to the Ma model using equations 11 and 12, the percentages and dispersions (approximating the dispersion of the long periods alone to be negligible), lie somewhere between the results from the 2 DM models. The KTG mass distribution which rises towards the low mass end, does, as expected, produce lower percentages of binaries identified and lower velocity dispersions.

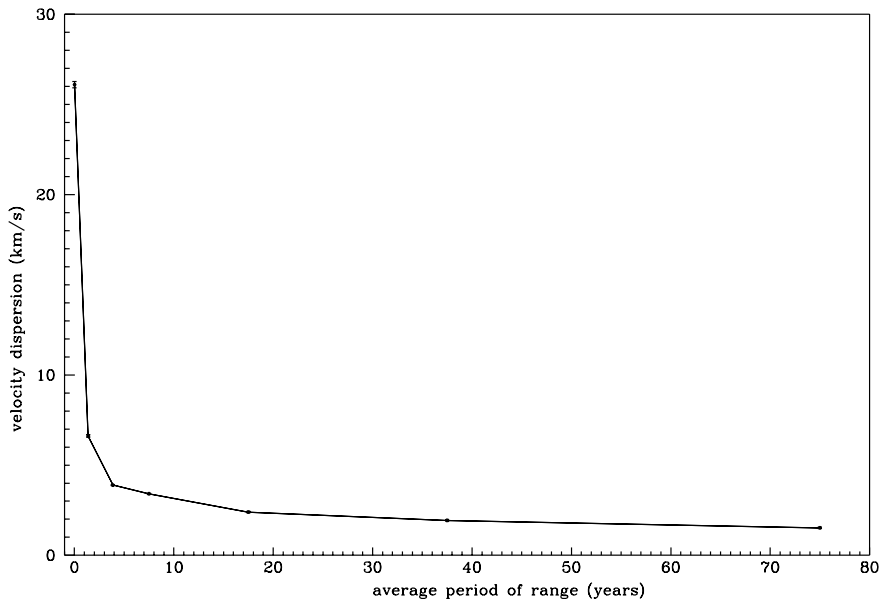

Figure 9: Variation of the velocity dispersion for different periods. Each point is positioned at the average period of the range for that simulation.

### 3.0.1 Models Which Produce Large Velocity Dispersions

Figure 9 shows the standard deviation of the velocity distribution of the DM10 model at different periods. The periods were chosen from a uniform distribution within a small range around the average value shown on the plot. It was not sensible to take single periods because the dispersion would be greater at integer multiples of the time between measurements (because if the large apastron velocities were measured once then they were measured every orbit): this does not represent the real life situation where observations are not taken at exact intervals. Only 1000 rather than 10000 stars were used for each of these simulations since, with such a limited period distribution, the velocity distribution was quicker to converge.

Periods of less than 5 years can produce large standard deviations of greater than 6 $\mathrm{km\,s^{-1}}$. When the calculation of Mateo et al. was repeated, without their error in the phase distribution (see Section 4), simulations were made for period ranges of 0.5 to 10 years, and 0.5 to 100 years. These produced dispersions of 5.5 and 4.3 km s$^{-1}$ respectively. This all implies that keeping the other parameters as before, only the situations where the binary fraction is close to 1.0 and the periods are almost all below 10 years, can the velocity dispersion produced by the binary stars be sufficient to account for that observed. For a threshold velocity of 4 km s$^{-1}$ (equivalent to a one sigma error of 0.9 km s$^{-1}$ on each velocity), one would expect to see more than 70% of the binaries in 10 years for periods below 10 years, requiring a far higher binary fraction observed than the 10–20% actually observed. This is true even allowing for the fact that not all these stars have been observed for 10 years.

Fixing the mass of the secondary to be equal to the mass of the primary also increases the standard deviation of the velocity distribution, the DM10 model producing a dispersion of over 6 km s$^{-1}$ (see Figure 10). Here, only 30% of the binaries would be identified in 10 years, for a threshold velocity of 4 km s$^{-1}$, which is much more in line with the observations.

As we have seen earlier, taking very high ellipticity orbits can produce high dispersions: however, these orbits are not physically possible and therefore not allowed using the cutoff radius method.

# 4 Main Differences Between the Simulations and Previous Work

### 4.0.2 Phase

Choosing the phase of an orbit randomly from a uniform distribution implies that the orbit has a constant angular velocity. This is only true for a circular orbit, as Kepler's 2nd law states that the line joining the 2 stars in an elliptical orbit sweeps out equal areas in equal times. The assumption of a constant angular velocity means that too many stars will be chosen in the part of the orbit close to apocentre where the true angular velocity is highest, relative to sampling equal time intervals throughout the orbit. It is around this part of the orbit that the stars have the largest orbital speed and this leads to an overestimate of the velocity dispersion caused by the binary stars.

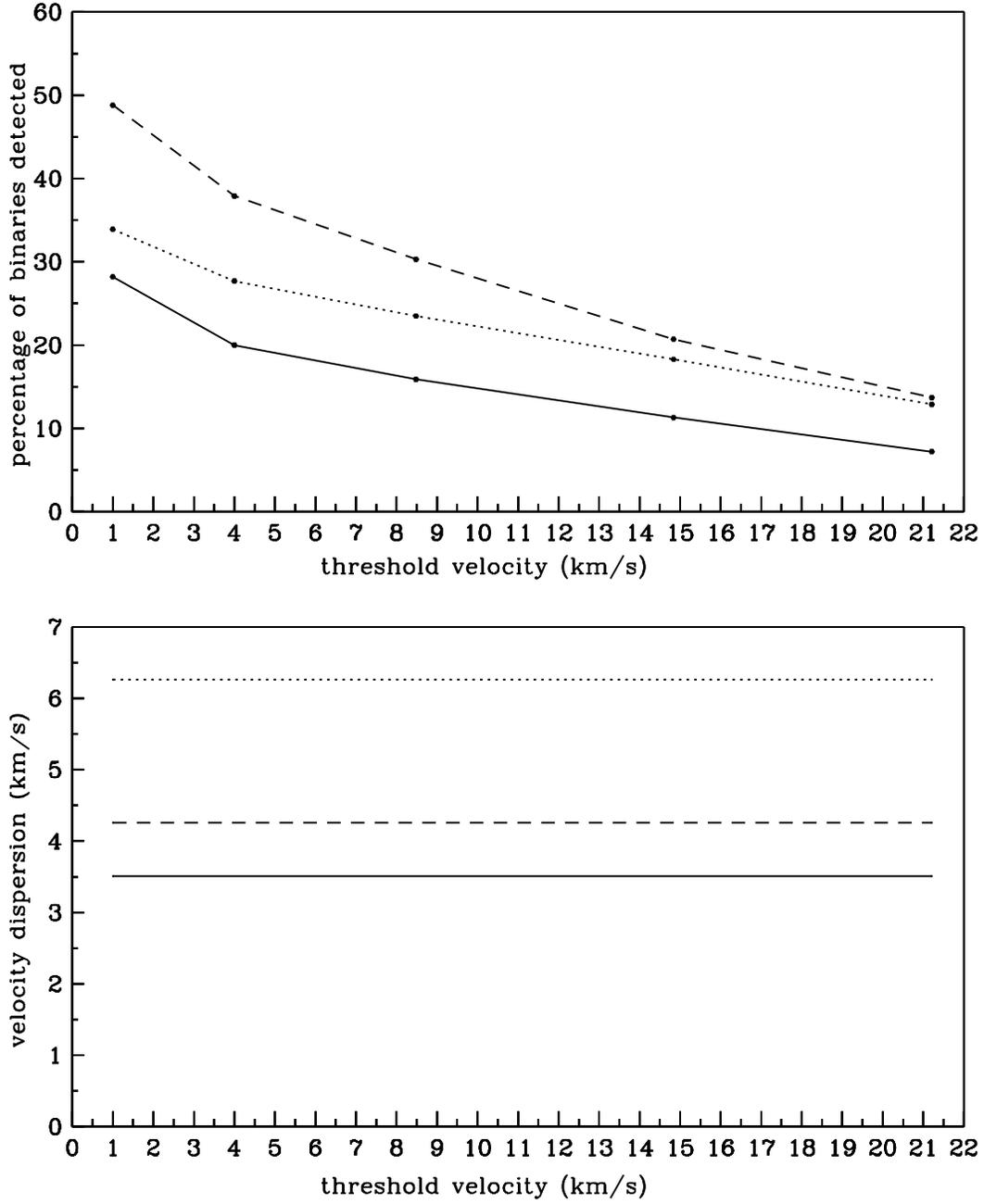

Figure 10: Ma, DM10 and DM30 models for M1=M2=0.8 R$_\odot$. These results are for 10 years of observations.

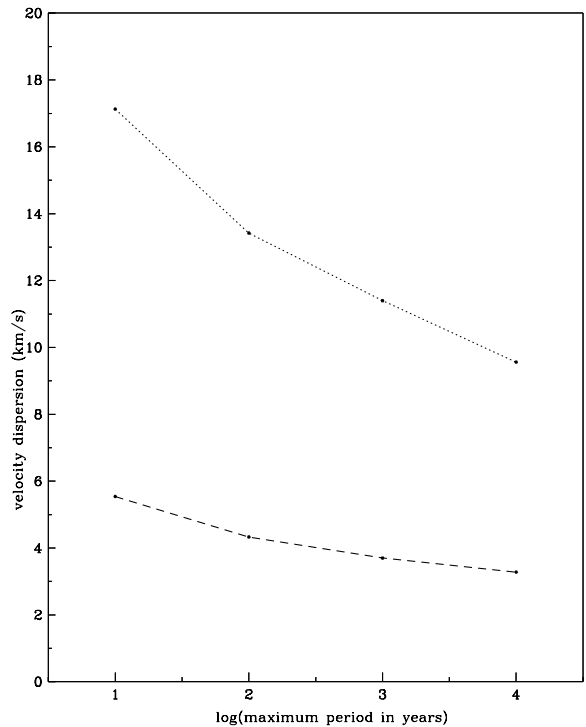

Figure 11: The velocity dispersion of the binary stars for the simulations published by Mateo et al., 1993 (dotted line) compared with the use here of their orbit distributions choosing the phase correctly (dashed line). The x-axis values are the upper cutoff period for the simulations, the lower cutoff in each case being 0.5 years

For the results produced by the other groups mentioned in Section 1.1, the overestimate is of the order of a factor of 3. Figure 11 shows the binary velocity dispersion (ie assuming 100 percent binary stars in the sample) obtained by Mateo et al. (1993) and the equivalent results choosing the orbital phase correctly using our model.

### 4.0.3 Binary Fraction

It is straightforward to calculate the size of the observed dispersion from the binary velocity dispersion, the binary fraction, and the intrinsic velocity dispersion.

Assuming the 2 dispersions both have a mean of zero,

$$\sigma_o^2 = (1-f)\sigma_i^2 + f\sigma_B^2, \tag{11}$$

where

$$\sigma_B^2 = \sigma_b^2 + \sigma_i^2. \tag{12}$$

Here $\sigma_o$ is the observed dispersion, $\sigma_i$ is the intrinsic velocity dispersion, $\sigma_b$ the calculated dispersion due to a binary fraction of 1.0, and $f$ is the binary fraction.

The difference between the standard deviation and biweight measurements noticed by Mateo et al. (1993) is an indication that the shape of the velocity distribution deviates from a Gaussian shape when binary stars are present. The fact that Mateo et al. chose the phase incorrectly suggests that this difference may not be as great for a more carefully chosen distribution. However, since the simulation reported in this paper only simulates the binary stars, calculating the effect of different binary fractions using the equations 11 and 12, this comparison has not been made. Instead, samples of small numbers of stars were drawn from the velocity distribution and K-S tests were performed to see at what level the distribution deviated significantly from a Gaussian shape.

### 4.0.4 Period and Radius Cutoffs

Previous simulations (for example Mateo et al. 1993) have taken an ellipticity range from a uniform distribution ranging from 0.5 to 1.0 and a period range with a lower bound of 0.5 or 1 year. Some cutoff value for the ellipticity must have been taken in previous simulations to avoid production of infinite velocities in the simulations, however this is not published and it is possible that unrealistic orbits with very high ellipticities were allowed. Figure 12 shows how the measured velocity dispersion varies for a fixed ellipticity. Here the period, taken from a distribution which is uniform in the logarithm of the period, ranges from 0.5 years to 10 years. As an alternative to this a minimum approach distance for the 2 stars in the binary has been taken, negating the need for artificial period and ellipticity cutoffs. When the same orbital distributions as those used by Mateo et al. were used a maximum ellipticity of 0.999 was taken (where ellipticity is $\sqrt{1-(b/a)^2}$ ), so as to avoid some of the unrealistic orbits shown by Figure 12. As can be seen from this figure and Figure 11, the effect is not sufficient to explain the difference between the simulations reported in this paper and the previously published results.

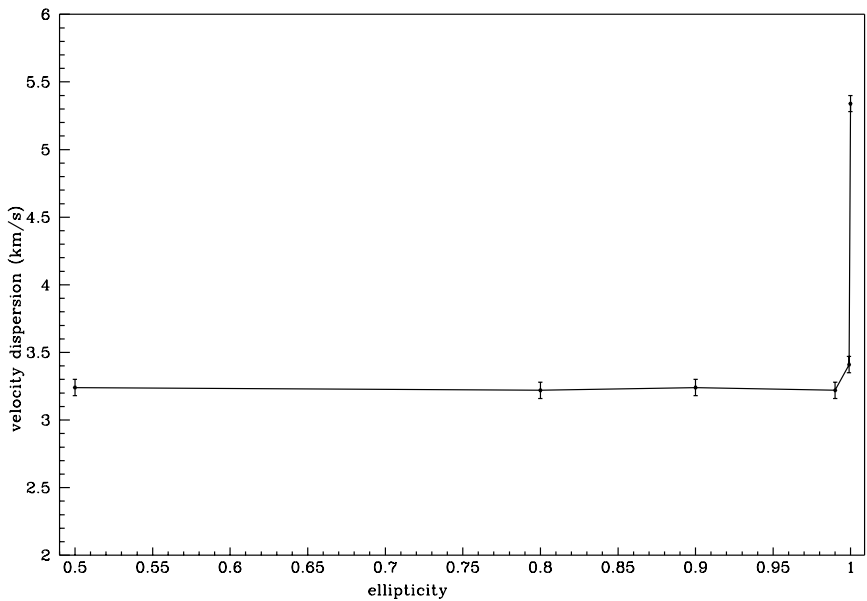

Figure 12: The velocity dispersion for the Ma model with periods between 0.5 and 10000 years, for different values of the ellipticity ($e = \sqrt{1 - (b/a)^2}$)

### 4.0.5 Number of Stars

Previous calculations have been aimed at simulating the exact experimental conditions of the observations. They therefore simulated velocities for only a small number of stars at a single epoch. The simulations were repeated of the order of 1000 times to obtain an average result for the velocity dispersion. A more robust and exact way to conduct the simulation is to use a much larger number of stars and calculate velocities all the way round the orbits.

# 5 Analysis of the Results

## 5.1 Comparison with the Observations

If the parameters defining binary star distributions in the Galaxy and dSph galaxies are the same, except for some loss of close orbits due to the expansion of the primary in the older population, then the effect of binaries on the measured velocity dispersion is small. If so, then there must be some other explanation for the large velocity dispersions which have been measured in dSph galaxies. When all the multi-epoch observations from dSph galaxies are added together, a binary fraction of between 0.1 and 0.16 is observed, with 4 to 12 years of measurements (Mateo 1994). Olszewski & Aaronson have between 5 and 10 roughly yearly measurements for their stars, and have detected a binary fraction of between 0.1 and 0.2. These results have not been fully published so there remain uncertainties about the exact answers, but they do claim a measuring error of about $1 \text{ km s}^{-1}$, so a threshold velocity of about $4 \text{ km s}^{-1}$ should be suitable for analysing their results. Duquennoy & Mayor found a binary fraction of 0.6 for the solar neighbourhood solar mass stars.

Considering the range of the results from the DM10 and DM30 models, between 12% and 24% of the binaries should be identified in 5 to 10 years, given a threshold velocity of $4 \text{ km s}^{-1}$ (see Figure 13). If 60% of the stars are binary stars, this means that we should actually identify 7% to 14% of the stars as binary stars. If the threshold velocity is $8.5 \text{ km s}^{-1}$ (equivalent to a $2 \text{ km s}^{-1}$ measuring error), we would expect between 4% and 10% of the binaries to be identified as such. The percentage of binaries that has been detected is a little on the high side (see the previous paragraph), but the discrepancy may well be caused by an underestimate of the measuring errors; for example, this could occur if there were broad wings on the error distribution. We correct for the fact that 28% of the binaries in DM30 and 15% in DM10 were rejected when compared with the DM1 model, where the radius of the primary was $1 \text{ R}_\odot$, because the minimum separation of the stars fell below the cutoff, and assume that these ex-binaries are still 'normal' stars in the sample. Then we conclude that the binary fraction should be 0.4 or 0.5, rather than 0.6. This results in a predicted percentage of binaries detected in 5–10 years, with a threshold velocity of $4 \text{ km s}^{-1}$, of 5% to 12%, which is more divergent from the observations. From the DM models, with a threshold velocity of $4 \text{ km s}^{-1}$, we should expect never to identify between 40% and 50% of the binary stars, however many years we observe for (Table 2). However, the dispersion caused by this fraction is very small, of the order of $1 \text{ km s}^{-1}$ (see Figures 5 and 6). From the shape of the velocity distribution formed by the binary orbit velocities alone, it appears that one would not reject the Gaussian distribution hypothesis with only 20 stars, but would in most cases with 40 stars. Therefore, with a binary fraction of 0.6, we would expect no divergence from a Gaussian shape to be detected at the level of our current observations.

Olszewski & Aaronson's observations of Draco have detected a binary fraction of 0.17 with up to 5 observations at roughly yearly intervals and measuring errors of about 1 km s$^{-1}$. This again is slightly on the high side.

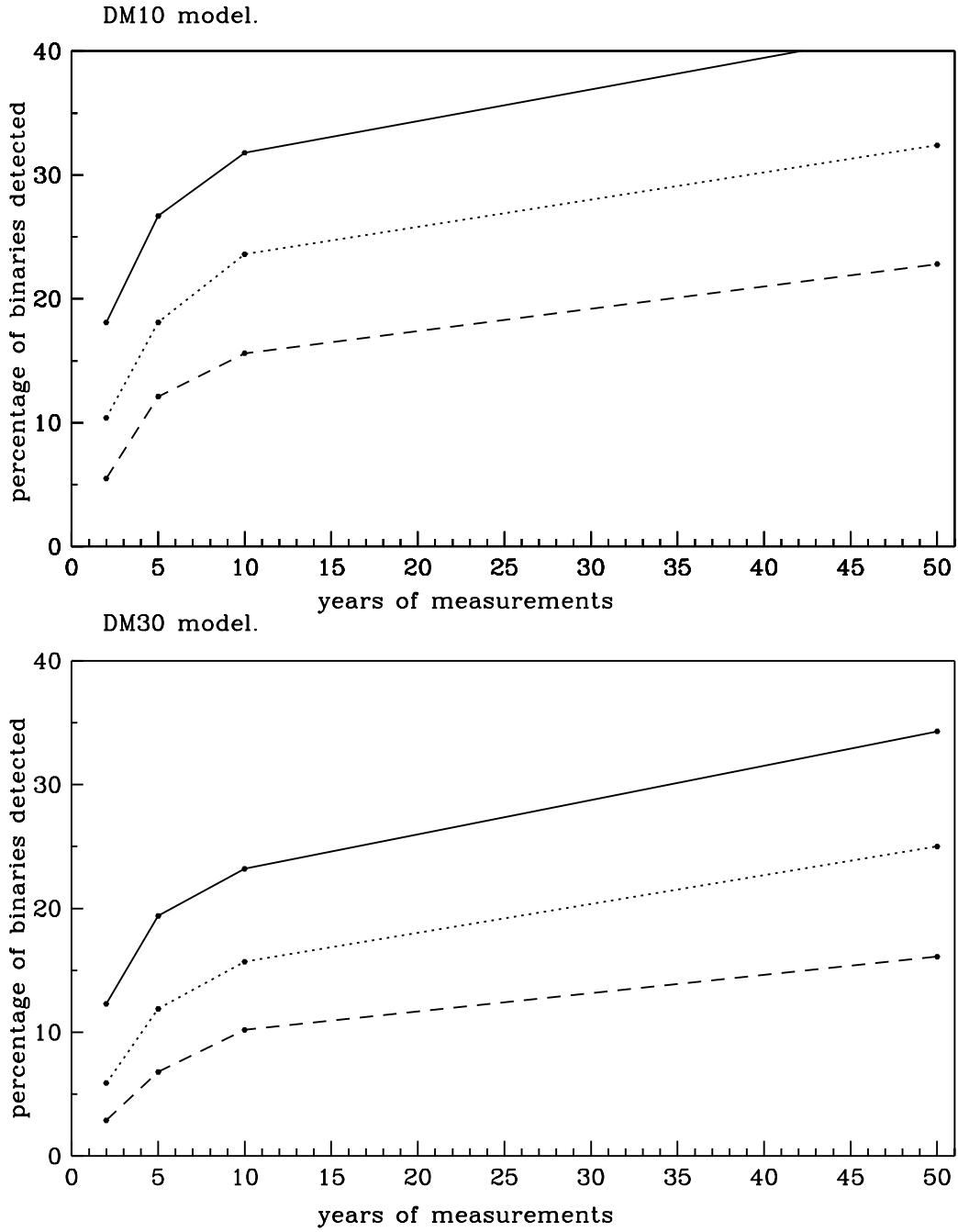

Figure 13: Variation of the percentage of binaries detected with number of years of observation. The top plot shows the results from the DM10 model with threshold velocities of 1, 4, and 8.5 km s$^{-1}$(solid, dotted and dashed lines respectively), and the lower plot presents results of the DM30 model adopting the same threshold velocities.

For the results by Hargreaves et al. (1994a, 1994b) 1 or 2 binaries out of 18 stars with multi-epoch observations in Sextans and Ursa Minor (6% to 11%) may have been found with 2 years of observation and a measuring error of 2 km s$^{-1}$. The DM models, and a binary fraction of 0.6, predict that 1.8% to 3.3% of the binaries should have been identified. We would, therefore, have expected to see 0 or 1 binary star. However, several of the multi-epoch measurements have more than 1 observation at each of the 2 epochs, as is the case for the strongest binary candidate. This leads to a considerably higher probability of identifying a binary star due to the effective decrease in measuring error resulting from the combination of several observations.

The results from the Ma model, restricting the period to various ranges (Figure 11) suggest that nearly all the binary stars which are identified within 10 years have periods of less than 100 years. This result was obtained by considering the sample with periods between 0.5 and 10000 year. Since we know that the distribution is uniform in the logarithm of the period, we can calculate the percentages of the total number of binary stars identified in the more restricted period ranges. The result is a negligible difference in the percentage between periods of 100 and 10000 years.

For the distribution to be such that the binary stars have a significant effect on the measurement of the velocity dispersion, we would require the sample to be biased towards lower periods and higher masses. In this case we would expect to detect between 30% and 70% of the binaries in 10 years of observation, assuming a threshold velocity of 4 km s$^{-1}$ (Table 2). Thus, the observations imply a binary fraction between 0.14 and 0.50. This requires a dispersion of about 10 km s$^{-1}$ or more for the binary stars to have a significant effect on the observed dispersion (Table 1). In these cases, the distributions of the orbital parameters of the binary stars are very different from the distributions of Galactic binary stars that have been detected through observations.

Simulations using a very small primary mass radius can also produce large dispersions: if the radius of the primary star were as small as 5 R$_\odot$, then dispersions greater than 5.5 km s$^{-1}$ could be produced, while stars of 1 R$_\odot$ can produce a dispersion of close to 10 km s$^{-1}$ (Figure 14).

The percentages of binary stars detected with the KTG models are only slightly less than those for the DM models. Thus, it is impossible, from these results, to rule out the KTG mass function, which rises towards lower masses. If anything it is slightly more in tune with the observations.

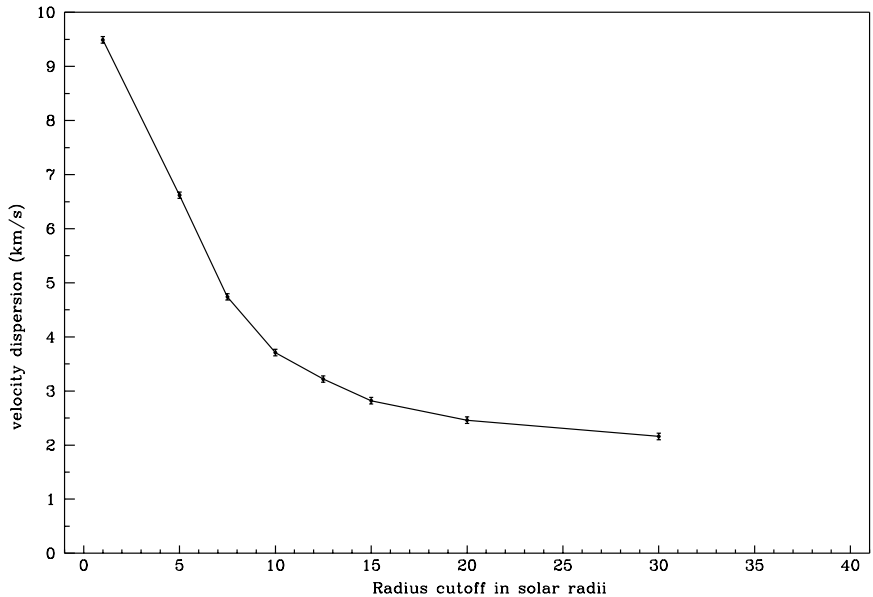

Figure 14: The velocity dispersion obtained from the DM models with different primary stellar radius cutoffs.

## 5.2 The Shape of the Velocity Distribution

The shape of the velocity distribution varies depending on the model chosen. Table 2 shows the sample size at which the distributions would probably be rejected at the three sigma level by a K-S test as being Gaussian. The distributions have broader wings than a Gaussian distribution. These sample sizes do not, however, reflect the number of stars that we would need to observe before recognising a component due to binary stars in an observed distribution. This is because the simulation has not taken into account the contribution to the distribution of the intrinsic velocity dispersion due to the mass of the galaxy.

A simple procedure was completed to illustrate the situation. Randomly chosen Gaussian deviates were added to the sample of velocities obtained from the binary star simulation. The intrinsic dispersion was chosen such that the total dispersion of the resulting sample was 6 $km\,s^{-1}$. K-S tests were then conducted for different sample sizes drawn from the new distribution.

Table 2: Results for the simulations using different models.

| Model | Comments | $\sigma_b$ km s$^{-1}$ | N | $V_T$ km s$^{-1}$ | $P_{10}$ | $P_N$ |
|---|---|---|---|---|---|---|
| Ma | 0.5–10 years | 5.5 | 100 | 4 | 75 | 1 |
|  | 1.0–10 years | 4.7 | 100 | 4 | 73 | 2 |
|  | 0.5–100 years | 4.3 | 100 | 4 | 56 | 3 |
|  | 0.5–1000 years | 3.7 | 80 | 4 | 39 | 8 |
|  | 0.5–10000 years | 3.3 | 60 | 4 | 31 | 17 |
| DM1 |  | 9.5 | 20 | 4 | 30 | 36 |
| DM5 |  | 5.6 | 20 | 4 | 27 | 38 |
| DM10 |  | 3.9 | 40 | 4 | 22 | 42 |
| DM20 |  | 2.5 | 40 | 4 | 18 | 47 |
| DM30 |  | 2.2 | 40 | 4 | 16 | 50 |
| DM40 |  | 1.9 | 60 | 4 | 13 | 53 |
| KTG10 |  | 2.9 | 40 | 4 | 48 | 49 |
| KTG30 |  | 1.7 | 40 | 8.5 | 19 | 79 |
| DM10 | 1–11 days | 26.5 | >200 | 4 | 98 | 0.4 |
|  | 11–1000 days | 6.6 | 180 | 4 | 84 | 5 |
|  | 2.7–5 years | 3.9 | 140 | 4 | 71 | 2 |
|  | 5–10 years | 3.2 | 100 | 4 | 69 | 5 |
|  | 10–25 years | 2.4 | 100–160 | 8.5 | 22 | 23 |
|  | 25–50 years | 1.9 | 100–160 | 8.5 | 9 | 31 |
|  | 50–100 years | 1.5 | 100–160 | 8.5 | 4 | 41 |
|  | 100–1000 years | 0.8 | 100–160 | 8.5 | 0.4 | 66 |
|  | 1000–10000 years | 0.4 | 100–160 | 8.5 | 0.01 | 87 |

Table 2 continued...

| Model | Comments | $\sigma_b$ km s$^{-1}$ | N | $V_T$ km s$^{-1}$ | $P_{10}$ | $P_N$ |
|---|---|---|---|---|---|---|
| Ma | e=0.9999 | 5.3 | 40–60 | 4 | 28 | 0 |
|  | e=0.999 | 3.4 | 40–60 | 4 | 28 | 1 |
|  | e=0.99 | 3.2 | 40–60 | 4 | 28 | 7 |
|  | e=0.9 | 3.2 | 40–60 | 4 | 32 | 31 |
|  | e=0.5 | 3.2 | 100 | 4 | 33 | 51 |
|  | e=0.0 | 3.2 | 100 | 4 | 33 | 54 |
| Ma | M2=M1 | 5.0 | 60 | 4 | 38 | 5 |
| DM10 | M2=M1 | 6.3 | 60 | 4 | 28 | 28 |
| DM30 | M2=M1 | 3.5 | 60 | 4 | 20 | 32 |

**Notes.** $\sigma_b$ is the standard deviation of the velocity distribution caused by the binary orbits obtained from the model defined in the first two columns.

N is the number of stars in the sample before a K-S test rejected the Gaussian hypothesis at the three sigma level. The samples were taken from the whole distribution in steps of twenty stars.

$V_T$ is the threshold velocity of the simulation for which the percentage $P_{10}$ of the binaries were detected in 10 years of yearly observations. $P_N$ is the percentage of the stars that would never be identified as binary stars because the velocity variations round the orbit are too small ever to be detected by the threshold velocity.

For the DM10 distribution 40 stars were sufficient to reject the Gaussian hypothesis. This distribution had a dispersion of 3.8 km s$^{-1}$ which was caused by binary star velocities alone. When we added a Gaussian deviate from an intrinsic dispersion of 4.6 km s$^{-1}$ to each star (making 6 km s$^{-1}$ in total, using equations 11 and 12), the K-S test did not reject the Gaussian hypothesis until samples contained as many as 1000 stars. The addition to each sample of 67% more stars, which had a velocity from the intrinsic dispersion, but no binary component, required the intrinsic dispersion to be 5.2 km s$^{-1}$. This sample, with a binary fraction of 0.6, required about 5000 stars before it was rejected at the three sigma level by the K-S test.

This result was compared with a simulation for which the dispersion caused by the binary stars alone is larger. In this case the DM10 simulation was performed, restricting the periods of the orbits to less than 3 years, the periods being drawn from a uniform distribution. The dispersion caused by only the binary stars was 6.2 km s$^{-1}$ and samples of about 220 stars were required before this distribution was rejected in the K-S test. We made the same calculation as described in the previous paragraph for this sample, but only conducted the experiment for a binary fraction of 0.6, because a binary fraction of 1.0 would require no contribution from an intrinsic dispersion. The intrinsic dispersion required by a binary fraction of 0.6 to make a total dispersion of 6 km s$^{-1}$ was 3.2 km s$^{-1}$. In this case about 500 stars were required for the Gaussian hypothesis to be rejected. Figure 15 shows these results for the 2 simulations.

In the light of these examples it seems unlikely that there should be any clear evidence for binary stars in the shape of the velocity distributions which have been obtained from dSph galaxies, because the largest sample sizes are about 80 stars. As an illustration we have combined the data from the observations reported by Hargreaves et al. (1994a, 1994b, 1995) for the Sextans Ursa Minor and Draco dSph galaxies. Each velocity distribution was normalised to a dispersion of 1 km s$^{-1}$ and then a K-S test was performed on the whole sample. This sample contained 74 stars and included the first epoch measurements from both suspected binary stars in Sextans. The probability from the K-S test was 0.8. Figure 16 shows the combined sample with the Gaussian function overlaid.

# 6  Conclusion

The velocity dispersion caused by binary stars with orbital parameters corresponding to the solar neighbourhood is small compared to the large velocity dispersions observed in dSph galaxies. The percentage of binary orbits that would be identified depends on the number of years of observation and on the precision of the velocity measurements. However, the simulations, which use orbital distributions derived from real observations, predict the identification of percentages of binary stars that are only slightly less than that actually observed.

To produce larger dispersions, more binary orbits with a mixture of lower periods, higher mass secondaries, or primaries with radii smaller than 10 R$_\odot$ are required. It is difficult to produce a velocity dispersion much above 6 km s$^{-1}$ without requiring restriction of the orbits to periods below about 5 years. For velocity dispersions from binary stars of the order of 6 km s$^{-1}$ to be significantly modifying the overall observed dispersion a binary fraction of close to 1.0 would be required. From observations of such a population spanning 10 years, around 30% of the binaries should be identified with such distributions. The observations are now slightly on the low side for this scenario, at 10%–20%, even accounting for the fact that not all these stars have been observed for 10 years.

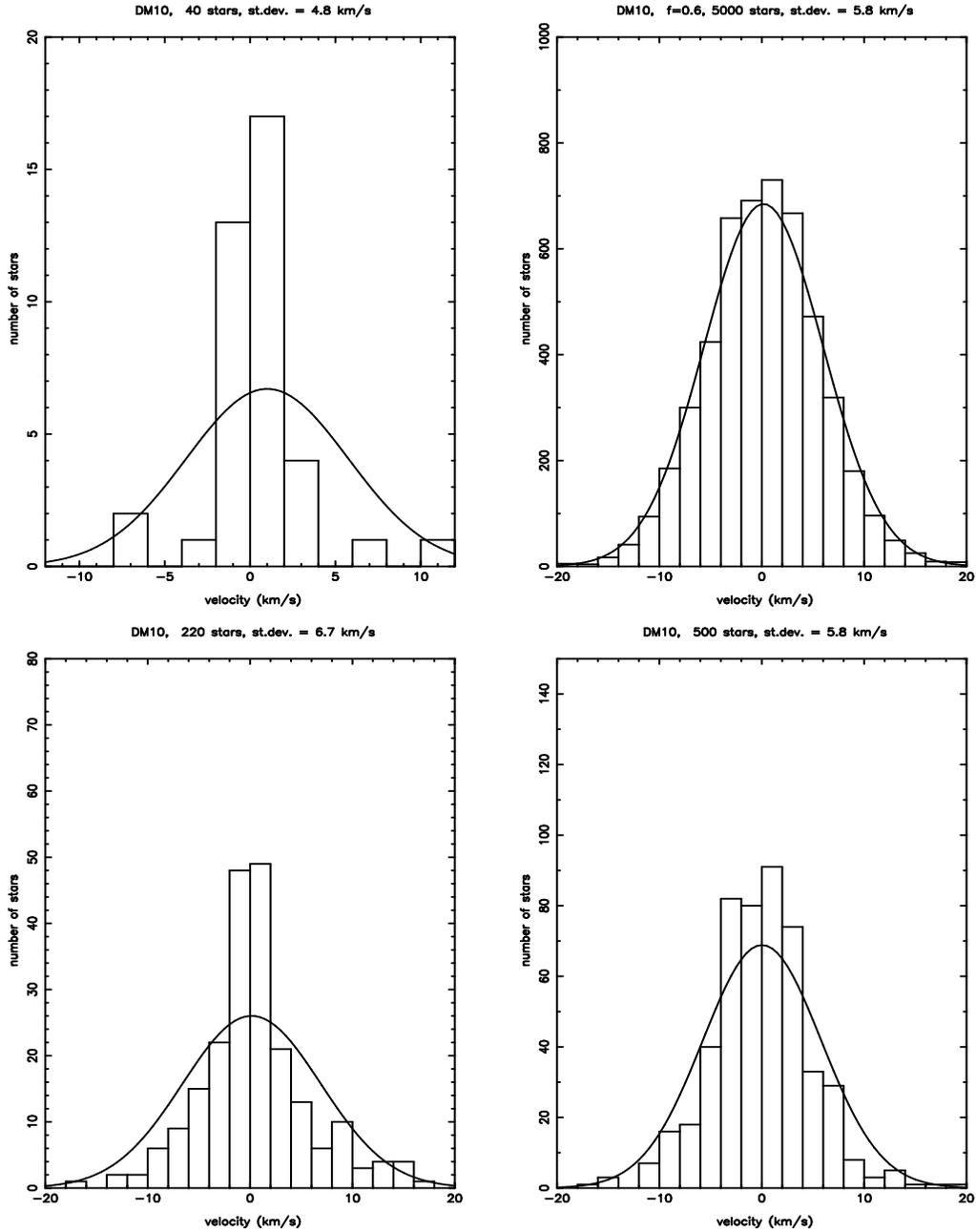

Figure 15: The plots on the left are the distributions for no intrinsic dispersion, and those on the right are for a binary fractions of 0.6. The upper plots show the standard DM10 model and the lower ones show the DM10 model with the periods restricted to below 3 years. The sample sizes are such that the K-S test would be very likely to reject the Gaussian hypothesis at the three sigma level. The Gaussians plotted have a one sigma width equal to the measured standard deviation (st. dev.) of the sample.

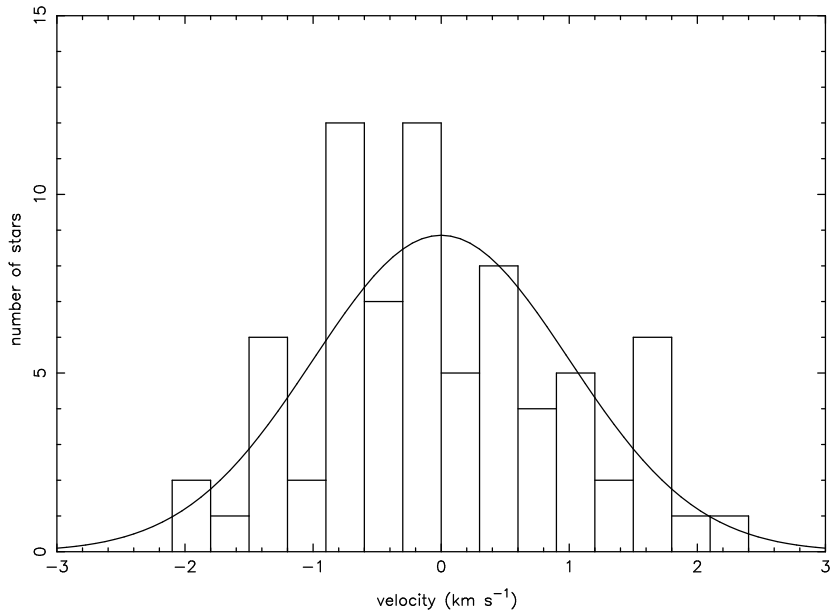

Figure 16: The combined sample of the Sextans, Ursa Minor and Draco data. The distribution from each sample has been normalised to a dispersion of 1 $\mathrm{km\,s^{-1}}$. The probability from the K-S test for the Gaussian distribution shown was 0.8.

At present it seems likely that some of the stars which observers have assumed to be binaries are erroneous detections, as a consequence of an underestimate of their measurement errors. Continuing high precision observations, when analysed using the results of this work, will be able to quantify the contribution of the binary stars to the observed velocity dispersion. Samples of greater than about 500 stars with single epoch measurements would be required to define the kinematic distribution function, but precise data for a few stars over a long time interval should be able to quantify the true significance of binary stars in the dynamics of dSph galaxies.

Table 1 Velocity dispersion required from the binary stars for binary orbits to account for the excess dispersion over that generated for a typical stellar mass-to-light ratio. All velocities are in km s$^{-1}$.

Measured dispersion = 7 km s$^{-1}$

| binary fraction | intrinsic dispersion 2 | 4 |
|---|---|---|
| 0.25 | 13.4 | 11.5 |
| 0.5  | 9.5  | 8.1  |
| 0.75 | 7.8  | 6.6  |
| 1.0  | 6.7  | 5.7  |

Measured dispersion = 10 km s$^{-1}$

| binary fraction | intrinsic dispersion 2 | 4 |
|---|---|---|
| 0.25 | 19.6 | 18.3 |
| 0.5  | 13.9 | 13.0 |
| 0.75 | 11.3 | 10.6 |
| 1.0  | 9.8  | 9.2  |

Table 2 Results for the simulations using different models.

| Model | Comments | $\sigma_b$ km s$^{-1}$ | N | $V_T$ km s$^{-1}$ | $P_{10}$ | $P_N$ |
|---|---|---|---|---|---|---|
| Ma | 0.5–10 years | 5.5 | 100 | 4 | 75 | 1 |
|  | 1.0–10 years | 4.7 | 100 | 4 | 73 | 2 |
|  | 0.5–100 years | 4.3 | 100 | 4 | 56 | 3 |
|  | 0.5–1000 years | 3.7 | 80 | 4 | 39 | 8 |
|  | 0.5–10000 years | 3.3 | 60 | 4 | 31 | 17 |
|  |  |  |  |  |  |  |
| DM1 |  | 9.5 | 20 | 4 | 30 | 36 |
| DM5 |  | 5.6 | 20 | 4 | 27 | 38 |
| DM10 |  | 3.9 | 40 | 4 | 22 | 42 |
| DM20 |  | 2.5 | 40 | 4 | 18 | 47 |
| DM30 |  | 2.2 | 40 | 4 | 16 | 50 |
| DM40 |  | 1.9 | 60 | 4 | 13 | 53 |
| KTG10 |  | 2.9 | 40 | 4 | 48 | 49 |
| KTG30 |  | 1.7 | 40 | 8.5 | 19 | 79 |
|  |  |  |  |  |  |  |
| DM10 | 1–11 days | 26.5 | >200 | 4 | 98 | 0.4 |
|  | 11–1000 days | 6.6 | 180 | 4 | 84 | 5 |
|  | 2.7–5 years | 3.9 | 140 | 4 | 71 | 2 |
|  | 5–10 years | 3.2 | 100 | 4 | 69 | 5 |
|  | 10–25 years | 2.4 | 100–160 | 8.5 | 22 | 23 |
|  | 25–50 years | 1.9 | 100–160 | 8.5 | 9 | 31 |
|  | 50–100 years | 1.5 | 100–160 | 8.5 | 4 | 41 |
|  | 100–1000 years | 0.8 | 100–160 | 8.5 | 0.4 | 66 |
|  | 1000–10000 years | 0.4 | 100–160 | 8.5 | 0.01 | 87 |
|  |  |  |  |  |  |  |
| Ma | e=0.9999 | 5.3 | 40–60 | 4 | 28 | 0 |
|  | e=0.999 | 3.4 | 40–60 | 4 | 28 | 1 |
|  | e=0.99 | 3.2 | 40–60 | 4 | 28 | 7 |
|  | e=0.9 | 3.2 | 40–60 | 4 | 32 | 31 |
|  | e=0.5 | 3.2 | 100 | 4 | 33 | 51 |
|  | e=0.0 | 3.2 | 100 | 4 | 33 | 54 |
|  |  |  |  |  |  |  |
| Ma | M2=M1 | 5.0 | 60 | 4 | 38 | 5 |
| DM10 | M2=M1 | 6.3 | 60 | 4 | 28 | 28 |
| DM30 | M2=M1 | 3.5 | 60 | 4 | 20 | 32 |

**Notes to Table 2.** $\sigma_b$ is the standard deviation of the velocity distribution caused by the binary orbits obtained from the model defined in the first 2 columns.

N is the number of stars in the sample before a K-S test rejected the Gaussian hypothesis at the three sigma level. The samples were taken from the whole distribution in steps of 20 stars.

$V_T$ is the threshold velocity of the simulation for which the percentage $P_{10}$ of the binaries were detected in 10 years of yearly observations. $P_N$ is the percentage of the stars that would never be identified as binary stars because the velocity variations round the orbit are too small ever to be detected by the threshold velocity.

# Figure Captions

**Figure 1.**

The velocity dispersion obtained for different primary masses for model DM10.

**Figure 2.**

Orbit of a binary star round the centre of mass.

**Figure 3.**

Geometry of the orbit of 2 binary stars

**Figure 4.**

Model Ma. The solid line in the bottom plot shows the velocity dispersion of the whole sample. The dashed and dotted lines show the percentage of binaries that would be detected (top plot) and the velocity dispersion of the residual sample (bottom plot) after 2 years of observations. The short dashes and long dashes show the same statistics for 10 and 20 years of observations respectively

**Figure 5.**

Model DM10. The solid line in the bottom plot shows the velocity dispersion of the whole sample. The dasked and dotted lines show the percentage of binaries that would be detected (top plot) and the velocity dispersion of the residual sample (bottom plot) after 2 years of observations. The short dashes and long dashes show the same statistics for 10 and 20 years of observations respectively

**Figure 6.**

Model DM30. The solid line in the bottom plot shows the velocity dispersion of the whole sample. The dashed and dotted lines show the percentage of binaries that would be detected (top plot) and the velocity dispersion of the residual sample (bottom plot) after 2 years of observations. The short dashes and long dashes show the same statistics for 10 and 20 years of observations respectively

**Figure 7.**

Model KTG10. The solid line in the bottom plot shows the velocity dispersion of the whole sample. The dashed and dotted lines show the percentage of binaries that would be detected (top plot) and the velocity dispersion of the residual sample (bottom plot) after 2 years of observations. The short dashes and long dashes show the same statistics for 10 and 20 years of observations respectively

**Figure 8.**

Model KTG30. The solid line in the bottom plot shows the velocity dispersion of the whole sample. The dashed and dotted lines show the percentage of binaries that would be detected (top plot) and the velocity dispersion of the residual sample (bottom plot) after 2 years of observations. The short dashes and long dashes show the same statistics for 10 and 20 years of observations respectively

**Figure 9.**

Variation of the velocity dispersion for different periods. Each point is positioned at the average period of the range for that simulation.

**Figure 10.**

Ma, DM10 and DM30 models for M1=M2=0.8 $R_\odot$. These results are for 10 years of observations.

**Figure 11.**

The velocity dispersion of the binary stars for the simulations published by Mateo et al., 1993 (dotted line) compared with the use here of their orbit distributions choosing the phase correctly (dashed line). The x-axis values are the upper cutoff period for the simulations, the lower cutoff in each case being 0.5 years

**Figure 12.**

The velocity dispersion for the Ma model with periods between 0.5 and 10000 years, for different values of the ellipticity ($e = \sqrt{1 - (b/a)^2}$)

**Figure 13.**

Variation of the percentage of binaries detected with number of years of observation. The top plot shows the results from the DM10 model with threshold velocities of 1, 4, and 8.5 km s$^{-1}$(solid, dotted and dashed lines respectively), and the lower plot presents results of the DM30 model adopting the same threshold velocities.

**Figure 14.**

The velocity dispersion obtained from the DM models with different primary stellar radius cutoffs.

**Figure 15.**

The plots on the left are the distributions for no intrinsic dispersion, and those on the right are for a binary fractions of 0.6. The upper plots show the standard DM10 model and the lower ones show the DM10 model with the periods restricted to below 3 years. The sample sizes are such that the K-S test would be very likely to reject the Gaussian hypothesis at the three sigma level. The Gaussians plotted have a one sigma width equal to the measured standard deviation (st. dev.) of the sample.

**Figure 16.**

The combined sample of the Sextans, Ursa Minor and Draco data. The distribution from each sample has been normalised to a dispersion of 1 $\mathrm{km\,s^{-1}}$. The probability from the K-S test for the Gaussian distribution shown was 0.8.